\documentclass[aps,twocolumn,nofootinbib,preprintnumbers,superscriptaddress,citeautoscript]{revtex4-1}
\pdfoutput=1
\usepackage{graphicx}
\usepackage{epsfig,amsmath,amsthm,amsfonts,mathrsfs,amssymb}
\usepackage[usenames,dvipsnames,table]{xcolor}
\usepackage{hyperref}
\usepackage[normalem]{ulem}
\allowdisplaybreaks

\DeclareRobustCommand{\SkipTocEntry}[5]{}

\hypersetup{
    pdfstartview={FitH},    
    pdftitle={},    
    pdfauthor={Massimiliano Rota, Sean J. Weinberg},     
    colorlinks=true,       
    linkcolor=blue,          
    citecolor=red,        
    filecolor=magenta,      
    urlcolor=blue           
}

\definecolor{green}{rgb}{0.1,0.8,0.2}
\definecolor{myblue}{rgb}{0.12,0.51,0.88}

\begin{document}

\title{Maximin is Not Enough}


\author{Massimiliano Rota}
\email{mrota@physics.ucsb.edu}
\affiliation{Department of Physics, University of California, Santa Barbara, CA 93106, USA}

\author{Sean J. Weinberg}
\email{sjasonw@physics.ucsb.edu}
\affiliation{Department of Physics, University of California, Santa Barbara, CA 93106, USA}

\begin{abstract}{
The RT formula for static spacetimes arising in the AdS/CFT correspondence satisfies inequalities that are not yet proven in the case of the HRT formula, which applies to general dynamical spacetimes.  Wall's maximin construction is the only known technique for extending inequalities of holographic entanglement entropy from the static to dynamical case.  We show that this method currently has no further utility when dealing with inequalities for five or fewer regions. Despite this negative result, we propose the validity of one new inequality for covariant holographic entanglement entropy for five regions.  This inequality, while not maximin provable, is much weaker than many of the inequalities satisfied by the RT formula and should therefore be easier to prove.  
If it is valid, then there is strong evidence that holographic entanglement entropy plays a role in general spacetimes including those that arise in cosmology. Our new inequality is obtained by the assumption that the HRT formula satisfies every known balanced inequality obeyed by the Shannon entropies of classical probability distributions.  This is a property that the RT formula has been shown to possess and which has been previously conjectured to hold for quantum mechanics in general. 
}
\end{abstract}

\maketitle
\tableofcontents

\section{Introduction}
\label{sec:intro}

The geometrical formula for entanglement entropy \cite{Ryu:2006bv}\cite{Hubeny:2007xt} in AdS/CFT \cite{Maldacena:1997re} reveals a great deal about the holographic structure of spacetime.   It provides, for example, a set of constraints in the form of linear entropy inequalities on the structure of CFT states with classical bulk duals. Studying these constraints provides important information about how bulk geometry is encoded in particular states of holographic CFTs.

The first such constraint to be discovered, the monogamy of mutual information inequality, was found in \citep{Hayden:2011ag}.   Additional constraints were then discovered in \citep{Bao:2015bfa} which introduced the notion of the \textit{holographic entropy cone}.  This concept plays a central role here so we review the construction in section \ref{sec:cones}.   (See also \cite{Marolf:2017shp}.)

The new entropy inequalities found in \citep{Bao:2015bfa} have a critical limitation: they have only been proven in the special case of static spacetimes. In this work we instead focus on entanglement constraints that apply
in generality.    The only technique currently available to prove that inequalities valid for static holographic entanglement entropy remain valid in the covariant case is 
 Wall's maximin technique \cite{Wall:2012uf}, which cannot be directly applied to the inequalities found in \citep{Bao:2015bfa}.  However, it is still a sensible question to ask whether 
there are other new inequalities, perhaps weaker than those of \citep{Bao:2015bfa}, which can be proved with the maximin method.  In section \ref{sec:implications} we demonstrate that this is not the case.

The underlying role that these inequalities play in quantum gravity is not clear at present.  While they are generally regarded as restrictions on the class of states that are suitable to have holographically dual spacetimes,
 we suggest here that they may also play an additional role: guaranteeing that holographic entanglement entropy is consistent with quantum mechanics.  This idea can be used to explore the possibility that
 covariant holographic entanglement entropy can be extended beyond the scope of the AdS/CFT correspondence \cite{Sanches:2016sxy}.  In particular, if the inequalities of holographic entanglement entropy
are strong enough, then extremal surface constructions in general realistic spacetimes must be consistent with quantum mechanics as explained in section \ref{sec:cones}.

%
%
%
%

\section{Holography and the Quantum Cone}
\label{sec:cones}


Consider a density matrix $\rho$ on a conformal field theory Hilbert space $\mathcal{H}_{\mathrm{CFT}}$.  We are particularly interested in $\rho$ if it can be regarded as a holographic state. By this we mean that $\rho$ can be viewed as a state on a semiclassical\footnote{We are interested here only in the limit where Newton's gravitational constant is sent to zero.} field theory with an asymptotically locally AdS spacetime background $M$.   Not just any density matrix on $\mathcal{H}_{\mathrm{CFT}}$ is a holographic state;  $\rho$ must satisfy certain criteria. 
In the following we will be interested in constraints associated with entanglement entropy.  

Consider a state $\rho$ on a particular time slice $\sigma$ on the conformal boundary $\partial M$, and on this time slice a subregion\footnote{By a subregion of $\sigma$, we mean a codimension 0 submanifold of $\sigma$ with boundary. 
 A subregion can be disconnected.} 
 $A \subset \sigma$.  The von Neumann entropy of $A$  in the state $\rho$, which we denote by $S(A)$, can be computed 
by finding the area of an HRT surface of $A$.
  The geometrical  HRT formula for $S(A)$ may imply new constraints on $\rho$ that are not satisfied by all quantum states.  
 
 In fact, if $M$ satisfies the null curvature condition, one such constraint is already known.  Suppose that $A$, $B$, and $C$ are three non-overlapping subregions of $\sigma$. Then, we must have \cite{Hayden:2011ag}\cite{Wall:2012uf}.
 \begin{equation}
 \label{eq:mmi}
 \begin{split}
 &S(AB) + S(AC) + S(BC) \\
  & -  S(A) - S(B) - S(C) - S(ABC) \geq 0
  \end{split}
 \end{equation}
 This inequality is called monogamy of mutual information (MMI) because it can be written in the form $I(A:B) + I(A:C) \leq I(A:BC)$.  
 
 Entropic constraints such as this are described by the \emph{holographic entropy cone} \cite{Bao:2015bfa} as we now explain.  Fix a positive integer $N$. We do not work with a CFT but instead directly consider some asymptotically locally AdS spacetime $M$ which may have multiple asymptotic regions and may even consist of multiple disconnected components.  Let $\sigma$ be a time slice of the conformal boundary 
 and suppose that $X_1, \ldots, X_N$ are (nonempty) subregions of $\sigma$ that do not overlap except,  perhaps, at their boundaries.  The regions $\{X_i\}$ actually define $2^{N}-1$ nonempty subregions of $\sigma$ by taking
 unions of the regions.  We will sometimes refer to these regions as subsystems.
 If $Z$ is one of these $2^N-1$ regions, let $S(Z)$ denote $1/4$ times the area of an HRT surface of $Z$ in Planck units.  The function $S$ can be conveniently regarded as an \emph{entropy vector}\footnote{We warn the reader that the usage of the word ``entropy'', although conventional, at this stage is not completely justified in the dynamical context. A vector of areas of extremal surfaces in fact could in principle be incompatible with the von Neumann entropies of any quantum state. This is not the case for static geometries \citep{Hayden:2016cfa}.} a point in ${\mathbf R}^{2^N-1}$
 whose components are the HRT entropies of each of the subsystems in some particular conventional order. We use the notation $S(X_I)_{I \in \mathcal{I} }$ for the vector associated with $S$.  Here, $\mathcal{I}$ is the
 power set of $\{1, 2, \ldots, N\}$ minus the empty set, and $X_I = \cup_{i\in I} X_i$.  
%
%

Now consider repeating the construction of $S(X_I)_{I \in \mathcal{I} }$ for every possible choice of $M$ satisfying the null curvature condition,\footnote{The null curvature condition is a reasonable energy condition to enforce in the classical limit as 
it arises in Einstein gravity as the $\hbar \to 0$ limit of the quantum null energy condition \cite{Bousso:2015mna}, which has been proven to hold in \cite{Balakrishnan:2017bjg}.}, every choice of $\sigma$, and every choice of $\{ X_i \}$.  The topological 
closure of the collection of all these entropy vectors is called the \emph{dynamical holographic entropy cone} for $N$ regions. It is indeed a convex cone in ${\mathbf R}^{2^N-1}$ and is thus defined by a (possibly infinite) collection
of linear inequalities.\footnote{In the static case the cone has been proven to be polyhedral in \citep{Bao:2015bfa}.} 
 
 An important modification of this cone construction is to proceed exactly as above except that we only consider asymptotically AdS spacetimes with the extra condition of being static.  
 We call the resulting cone the \emph{static holographic entropy cone} for $N$ regions.  This was simply referred to as the ``holographic entropy cone'' in \cite{Bao:2015bfa}.  However, we avoid this terminology to 
 stress the importance of the open problem to determine whether or not the static and dynamical entropy cones are equal to each other.  For $N \leq 4$, the static and dynamic cones are exactly known and are in fact identical.
 For $N \geq 5$, neither the static nor dynamical cones are known and there are several inequalities that have been proven for the static cone \cite{Bao:2015bfa} but thus far have no general proof for the dynamical case---see the appendix \ref{sec:appendix}.
 
 The only inequalities that are known for the dynamical cone are: positivity, subadditivity (SA), Araki-Lieb (AL), strong subadditivity (SSA), weak monotonicity (WM) and MMI.  
  This does not mean that these are the only possible inequalities,
 but instead suggests the difficulty of extending arguments for holographic entanglement inequalities from the static to general case.  In fact, we prove in section \ref{sec:implications}
 that the only known method for generalizing static inequalities, Wall's ``maximin'' technique, cannot currently be used to obtain any further dynamical inequalities.

It is useful to give a name to the cone defined by known dynamical inequalities alone.  Thus, for $N\geq3$,  we define the \emph{MMI cone} for $N$ regions as the convex cone
 in ${\mathbf R}^{2^N-1}$ defined by all possible realizations of positivity, subadditivity, Araki-Lieb and MMI \footnote{As a consequence of MMI, strong subadditivity and weak monotonicity are redundant and there is no reason to include them in the definition.}.  The dynamical holographic entropy cone is a subcone of the MMI cone.
 We also note that the MMI cone is is precisely equal to both holographic entropy cones for $N=3$ and $N=4$ \cite{Bao:2015bfa}.
 
\addtocontents{toc}{\SkipTocEntry}
\subsubsection*{Holography in General Spacetimes and New Dynamical Inequalities}

Many of the most important spacetimes in physics, like those that arise in cosmology,
are beyond the scope of the AdS/CFT correspondence. 
Despite this, properties of AdS/CFT may apply to  quantum gravity broadly.
Holographic entanglement is one such property, and its extension has the
capacity to betray aspects of quantum states for general spacetimes.

A construction analogous to holographic
entanglement entropy, but which arises in generic realistic\footnote{By a realistic spacetime we mean one satisfying conditions given in \cite{Bousso:2015qqa}
 which include the null curvature condition} spacetimes, was introduced 
in \cite{Sanches:2016sxy}.   The basic idea of this approach is simple: in spacetimes
 with no conformal boundary,
 we consider instead a surface which has properties suggestive of holographic
 duality and which reduces to the conformal boundary
 in the appropriate cases.  Then, on such a surface, we compute the area of
 anchored extremal surfaces.   Holographic screens \cite{Bousso:1999cb}, and especially their special case of  past and future holographic
 screens \cite{Bousso:2015mqa}\cite{Bousso:2015qqa}, are ideal surfaces for these purposes.\footnote{We focus on the screen entanglement proposal
 because it is a generalization of the HRT formula that has many desirable features \cite{Nomura:2016aww}\cite{Nomura:2016ikr}.  Our considerations here apply, however, to any reasonable
 extension of covariant holographic entanglement entropy to general spacetimes like that of \cite{Miyaji:2015yva}.  Our primary concern is whether or not
 any such extension is consistent with quantum mechanics.}
 
 The   \emph{screen entanglement proposal} \cite{Sanches:2016sxy}
 is the postulate that minimal extremal surfaces anchored to the unique time slices of past or future holographic screens are
  in fact von Neumann entropies for a quantum state despite the fact that no framework is known
for the quantum mechanics of completely general spacetimes.  
This proposal is falsifiable  even without such a framework as we now explain.
 Let $\{X_i \}_{i=1}^N$ be subregions of  one of the unique time slices of a past holographic screen
 and, following this extremal surface construction, compute the entropy\footnote{As in the dynamical setting in AdS/CFT, here we use the word ``entropy'' in a loose sense.} vector $S(X_I)_{I \in \mathcal{I} } \in {\bold R}^{2^N-1}$.
 If any entropy vector obtained in this way lies outside of the quantum entropy cone\footnote{
 The quantum entropy cone \cite{Pippenger:2003aa} for $N$ parties is defined as follows. 
 Consider a density matrix on the tensor product of  $N$ Hilbert spaces.  An entropy vector can
then be obtained by computing the von Neumann entropy of every subsystem.  
The quantum entropy cone is the region of ${\mathbf R }^{2^N-1}$ spanned 
by these vectors for all choices of the Hilbert spaces and every density matrix. }
for $N$ parties, then the screen entanglement proposal cannot be valid.

There is evidence already that the proposal will never be falsified in this way.  The exact inequalities known to be satisfied by the dynamical
holographic entropy cone are the inequalities also proven for entropy vectors obtained from holographic screens  \cite{Sanches:2016sxy}. 
In particular, the screen generalization of the dynamical holographic entropy cone is a subcone of the MMI cone which includes all known inequalities
of the quantum cone.  
 However, as discussed in section \ref{sec:inequalities}, there are many further inequalities that are conjecturally
valid for quantum states.  We are thus led to the following nontrivial conjecture: \emph{the dynamical holographic entropy cone for $N$ regions, as well as its generalization
for holographic screens, is a subcone of the quantum entropy cone for $N$ parties}. This conjecture is known to be valid for $N \leq 4$, but the $N=5$ case is extremely difficult.\footnote{In the static case, this is known to be true for all $N$ \citep{Hayden:2016cfa}.}

The power of this conjecture is startling: it guarantees that the screen entanglement proposal is consistent with quantum mechanics in the following sense. If $N$
is taken to be large enough, the regions $\{X_i\}$ can cover all of a time slice of a screen while at the same time having area around the UV cut-off scale.  The validity 
of our conjecture would guarantee the existence of a single quantum state on a tensor product of $N$ Hilbert spaces, each assigned to a local subregion $X_i$, whose von Neumann
entropies exactly reproduce all possible entropy computed with the geometrical proposal involving extremal surfaces.

This conjecture is a major reason that we strongly advocate the search for inequalities satisfied by the dynamical holographic entropy cone. (Such inequalities would most
likely also apply to past holographic screens by virtue of their convexity properties.)  
In principle such inequalities could be used to prove the conjecture directly: if the dynamical entropy cone is bounded by a polyhedral cone, then we can attempt to find
quantum states  with von Neumann entropies that exactly match the extremal rays of the bounding cone.

%
%

Such an approach is daunting.  Fortunately, in section \ref{sec:inequalities} we suggest an alternative starting point that is likely to be much easier and
already has the potential to either strongly support or oppose the screen entanglement construction.  Rather than
searching for inequalities of the dynamical cone and then performing the difficult task of finding quantum states that reproduce corresponding extremal rays,
we advocate the search for a proof that the dynamical cone satisfies a single additional inequality that is given in equation \eqref{eq:purified_zhang}.  This inequality is much weaker
than those known to hold for the static cone so it should be easier to prove.  On the other hand, as explained below, \eqref{eq:purified_zhang} is a nontrivial inequality for five parties
that is expected to hold for the quantum cone.  If it were proven for the dynamical entropy cone then the screen entanglement proposal would stand on more firm ground.  If it were
violated by any ray of the dynamical entropy cone, then holographic entanglement entropy is unlikely to generalize beyond AdS/CFT.

\section{Quantum and Classical Entropy Inequalities}
\label{sec:inequalities}

Given the considerations above, it is of significant interest in quantum gravity to study the inequalities satisfied by the von Neumann entropy.
For $N \geq 3$, there is no proven unconstrained\footnote{See \citep{Cadney:2011aa} for a definition and analysis of constrained inequalities.}
inequality other than strong subadditivity.
However, \cite{Cadney:2011aa} suggested a possibility: that every balanced \cite{Chan:2003aa} inequality satisfied by the Shannon entropies of the marginals of probability distributions on $N$ variables is also a valid inequality for the von Neumann entropies of subsystems of $N$-partite density matrices.
In the following we will assume that this conjecture is true and explore its consequences for the holographic cones introduced in the previous section.

The study of classical inequalities in the holographic context is also motivated by the fact that for any $N$ the static holographic entropy cone is inside of the stabilizer cone \cite{Hayden:2016cfa}, and that stabilizer states satisfy all balanced classical inequalities \cite{Gross:2013aa}. We will briefly comment in the discussion about the possibility that the conjecture \cite{Cadney:2011aa} is false.

Many known balanced classical inequalities can easily be shown to follow from known inequalities of the dynamical holographic entropy cone, even when extended to general spacetimes with the approach suggested in section \ref{sec:cones}.
For $N=4$ parties, infinitely many classical inequalities are known \cite{Matus:2007aa}.  It is straightforward to check that all of these inequalities are satisfied by the dynamical holographic entropy cone as a consequence of MMI.
This must be the case:  MMI implies the Ingleton inequality and, for $N=4$, the cone of stabilizer states is exactly given by the Ingleton inequality along with SA, AL, SSA and WM \cite{Linden:2013aa}.
Similarly, the MMI cone implies all of the families of inequalities found in \cite{Matus:2007aa}, which apply to the case of $N=5$, as well as the families of \cite{Zhang:1998aa} and \cite{K.-Makarychev:2002aa}, which provide inequalities for every $N \geq 5$. 

However, a further generalization of \cite{K.-Makarychev:2002aa}  was found in \cite{Zhang:2003aa} that gives an inequality for any $N=k+4\geq 6$ \footnote{In the $k=1$ case the inequality is actually trivial, since it is simply implied by SA and SSA.} random variables $\{A_1, \ldots, A_k, C, D,E,F \}$:
\begin{equation}
\label{eq:zhang}
\begin{split}
&\sum_{i=1}^k(S(A_iCF)+S(A_iDF))-\sum_{i=1}^kS(A_iCDF)\\
&-S(A_1...A_kF)-(k+1)S(CDEF)-kS(CF)\\
&-kS(DF)-(k-1)S(EF)+(k+1)S(CDF)\\
&+kS(CEF)+kS(DEF)\geq 0.
\end{split}
\end{equation}
For $k=2$ this furnishes an inequality that is not implied by MMI.  It is, however, implied by \eqref{eq:new_inequality4} which is a known inequality of the static holographic entropy cone \cite{Bao:2015bfa}. 

%
%

Given a balanced inequality for $N$ parties, either classical or quantum, one can obtain new inequalities for fewer variables by assuming that the probability distribution, or density matrix, factorizes. Applying this procedure to \eqref{eq:zhang}, one can only get other members of the family (for fewer variables) or new inequalities which are implied by the MMI cone. However, under the assumption that \eqref{eq:zhang} are valid quantum inequalities, there are more possibilities.

Introducing the purification $O$ of $\{A_1, \ldots, A_k, C, D,E,F \}$, new inequalities, which are physically equivalent to \eqref{eq:zhang}, can be obtained using the standard procedure that gives WM from SSA.\footnote{To derive a new inequality choose one of the variables and replace each term where the variable appears in the inequality with the entropy of the complementary subsystem.} Furthermore, for each of these inequalities, one can reduce the number of variables assuming the factorization of the density matrix as described above. 

We have implemented this procedure for \eqref{eq:zhang} in the case $k=2$, for all possible choices of variables, to obtain new $N=5$ inequalities. Most of the new inequalities are trivial or implied by the MMI cone. However, one is new:
\begin{equation}
\label{eq:purified_zhang}
\begin{split}
&2S(ABC)+2S(ABD)+3S(ABE)+S(ACE)\\
&+S(ADE)+S(BCE)+S(BDE)-3S(AB)-S(AE)\\
&-S(BE)-S(CDE)-S(ABCD)-2S(ABCE)\\
&-2S(ABDE)\geq 0.
\end{split}
\end{equation}

This inequality, like any balanced classical inequality, is implied by the static holographic entropy cone.  It is interesting to note, however, that it is in some sense ``weak.''  Suppose that we
consider the cone defined by SA, AL, MMI, and some subset $I$ of the known inequalities of the static entropy cone \cite{Bao:2015bfa} which are listed in the appendix.    \eqref{eq:purified_zhang}  is implied by our cone as long as
$I$ contains any one of the inequalities in the appendix with the exception of \eqref{eq:new_inequality5}.  Compare this to the $N=6$ case of  \eqref{eq:zhang} which is not implied unless $I$ contains \eqref{eq:new_inequality4}.
The weakness of \eqref{eq:purified_zhang} is a desirable feature: this new inequality may be significantly simpler to prove for the dynamical holographic entropy cone than the stronger inequalities that appear in the appendix.

%

In dynamical situations, it is currently not know if the inequalities of \cite{Bao:2015bfa} are valid. On the other hand, under the assumption that \eqref{eq:zhang} is a true quantum inequality, consistency with quantum mechanics would require that \eqref{eq:purified_zhang} is always satisfied in AdS/CFT. A sensible question then is whether one can bound the dynamical holographic entropy cone by new inequalities which are strictly weaker than the ones of \cite{Bao:2015bfa}, but still strong enough such that \eqref{eq:purified_zhang} is guaranteed to be satisfied. Furthermore, if these inequalities could be proved using the technique of maximin, they would be true also in the context of holographic screens, providing evidence for the conjecture discussed in section \ref{sec:cones}. The search for these inequalities is the focus of the next section.

\section{Maximin Provable Inequalities}
\label{sec:implications}

Motivated by the observations discussed above, we would like to find new inequalities for $N=5$ that can be proved in the dynamical setting using the techniques of maximin.  For an inequality to furnish an interesting bound on the dynamical holographic entropy cone, it must be independent from the inequalities that define the MMI cone.  Such new inequalities could even be sufficient to imply \eqref{eq:purified_zhang}.  Unfortunately, the remainder of this section demonstrates
 that no such maximin provable bound on the dynamical cone can be obtained given the currently known inequalities of the static cone; the most stringent inequality that can be proved in the dynamical setting is in fact MMI.

Consider an inequality which has been proven in the static case, and write it in the following form 
\begin{equation}
\sum_{I\in\mathcal{I^+}}c_I^{(+)}S(X_I)-\sum_{I\in\mathcal{I^-}}c_I^{(-)}S(X_I)\geq 0
\label{eq:inequality_form}
\end{equation}
where $\mathcal{I}^{\pm}$ are chosen such that $\mathcal{I}^+\cup\mathcal{I}^-=\mathcal{I}$ and all the $c_I^{\pm}$ are non-negative. To be able to extend the proof to dynamical situations using the techniques of maximin, it is necessary that the negative terms do not overlap.  That is, if $I,J\in\mathcal{I^-}$, then we must have
\begin{equation*}
X_I\subset X_J\;\; \text{or}\;\; X_J\subset X_I\;\; \text{or}\;\; X_I\cap X_J=\emptyset.
\label{eq:DMP_condition}
\end{equation*}
Any inequality which has been proven in the static case and satisfies this condition will be said to be \textit{directly maximin provable} (DMP) because the maximin procedure immediately implies its validity for the dynamical holographic
entropy cone. Note however that any conical combination of DMP inequalities is obviously also a true inequality in a dynamical situation, although it need not have the DMP form.


We can list all possible choices of $\mathcal{I}^-$ which are in DMP form for $N=5$.    The possible (overlapping) cases up to permutations are
\begin{equation}
\label{eq:maximin_structure}
\begin{split}
&\mathcal{I}^-\subseteq \mathcal{I}_1^- \equiv \{ABCD,ABC,AB,A,B,C,D,E,ABCDE\}\\
&\mathcal{I}^-\subseteq \mathcal{I}_2^- \equiv \{ABCD,AB,CD,A,B,C,D,E,ABCDE\}\\
&\mathcal{I}^-\subseteq \mathcal{I}_3^- \equiv \{ABC,DE,AB,A,B,C,D,E,ABCDE\}.
\end{split}
\end{equation}
Here we have chosen to label our subsystems so that  $\mathcal{I}=\mathscr{P}(\{A,B,C,D,E\})\backslash\{\emptyset\}$.

To look for new DMP inequalities, it is more convenient to work with dual cones. For our purposes this simply means that for a given non-redundant  inequality $\sum_{I\in\mathcal{I}} c_I S(X_I) \geq 0$ of a cone, we consider the list $(c_I)_{I\in\mathcal{I}}$ of its coefficients and we treat that list as an extremal ray of a new cone in ${\mathbf R}^{2^N-1}$.  Any redundant inequality for an original cone gives a ray in the corresponding dual cone which is not extremal.  
For this reason, if $S$ and $T$ are two convex cones with $S \subset T$,
then the dual cones $S^*$ and $T^*$ satisfy $T^* \subset S^*$

Consider now the cone $\Gamma_5$ obtained by  starting with the MMI cone (which we denote by $\Omega_5$ from now on) and cutting it with the inequalities of \citep{Bao:2015bfa} listed in the appendix. 
The static holographic entropy cone with $N=5$ is a subset of $\Gamma_5$ which is itself a subset of $\Omega_5$.
In the dual picture, $\Omega_5^*$ is a subset of $\Gamma_5^*$, which is a subset of the dual to the static holographic entropy cone. 
  

Consider an arbitrary subset  $\mathbb{G}=\{g_1, \ldots , g_k \}$ of the extremal rays of $\Gamma_5^*$. We denote the subcone of $\Gamma_5^*$ that they generate as $C_{\mathbb{G}}$. Any element $r$ of $C_{\mathbb{G}}$ is a conical combination of the generators:
\begin{equation}
r=\sum_{i=1}^k\alpha_ig_i 
\label{eq:conical_combination}
\end{equation}
where $\alpha_i \geq 0$ for each $i$.

 A point in $r$ in $C_{\mathbb{G}}$ is dual to a (potentially redundant) inequality for $\Gamma_5$.   We want to know whether or not $r$ has
 one of the structures of \eqref{eq:maximin_structure}.  That is, we wish to know if $r$ corresponds to an inequality of the form \eqref{eq:inequality_form}
 such that $\mathcal{I}^- \subset \mathcal{I}_s^-$ for some $s \in \{1,2,3 \}$.  Define $\mathcal{I}_s^+ =\mathcal{I} \setminus \mathcal{I}_s^-$.  If
 we write $r = (r_I)_{I\in\mathcal{I}}$, then the desired condition on $r$ is 
\begin{equation}
r_I\geq 0\, \qquad \forall \: I\in\mathcal{I}_s^+
\label{eq:DMP_condition}
\end{equation}
Note that we do not impose any constraint on the components $r_I$ with $I\in\mathcal{I}_s^-$.

 Now we consider the coefficients in equation \eqref{eq:conical_combination} which can be regarded as a vector $\alpha = (\alpha_i)_{i=1}^k$.  Choose a particular DMP structure
 by taking $s \in \{1,2,3\}$.  Then, \eqref{eq:DMP_condition}  gives a collection of inequalities we can impose on $\alpha$.   The fact that \eqref{eq:conical_combination} must be a conical combination 
 means that we have $\alpha_i\geq 0$ for all $i$.  This set of inequalities (both those of \eqref{eq:DMP_condition} and the positivity conditions) defines a new cone in $k$ dimensions so we can find its extremal rays. These can then be mapped to a collection of rays in $C_{\mathbb{G}}$
 which furnishes a subcone $C_{\mathbb{G}}^{\mathcal{I}_s^-} \subseteq C_{\mathbb{G}}$ of DMP rays for a particular structure $\mathcal{I}_s^-$. 
 
 We can repeat this process not only for $\{\mathcal{I}_s^- \:|\: s=1,2,3\}$,
  but also with the larger collection $P$ of all permutations of these three structures.  This gives a family of cones 
   $\{C_{\mathbb{G}}^{\mathcal{I}^-} \: | \: \mathcal{I}^-  \in P \}$ inside $C_{\mathbb{G}}$. The space spanned by all rays (i.e. inequalities in the original picture) in $C_{\mathbb{G}}$ that can be proved (both directly and indirectly) for dynamical situations using maximin techniques is then
\begin{equation}
C_{\mathbb{G}}^{\text{Mm}}=\text{cone}\left(\bigcup_{\;\mathcal{I}^- \in P }C_{\mathbb{G}}^{\mathcal{I}^-}\right),
\end{equation}
the conical hull of the cones in $\{C_{\mathbb{G}}^{\mathcal{I}^-} \: | \: \mathcal{I}^-  \in P \}$.

In principle we could obtain all the DMP inequalities of interest by applying the previous procedure  with $\mathbb{G}$ taken to be the full set of generators of $\Gamma_5^*$, rather than to a smaller subset. However this is impractical due to the highly complicated structure of the cone. Furthermore, we are only interested in the complement of $\Omega_5^*$ in $\Gamma_5^*$.   

Consider a facet $f$ of $\Omega_5^*$.  Since we are working in the dual picture, this is one of the extremal rays of $\Omega_5$ (see the appendix \ref{sec:appendix}). The hyperplane $\Sigma_f$ corresponding to the facet divides the full space into two halfspaces: $R^{2^N-1}=H_f^{(+)}\cup H_f^{(-)}\cup\Sigma_f$. The closure of one of the halfspaces, which we choose to be $H_f^{(+)}$, contains the whole $\Omega_5^*$. We can then cut $\Gamma_5^*$ using $\Sigma_f$ and focus only on the subcone in $H_f^{(-)}$. We classify the extremal rays of $\Gamma_5^*$ according to their location. We denote by $\mathfrak{g}^+$ the collection of generators of $\Gamma_5^*$ in $H_f^{(+)}$,  by $\mathfrak{g}^-$ the generators in $H_f^{(-)}$, and by $\mathfrak{g}^0$ the generators that lie on $\Sigma_f$. 

One could hope that the extremal rays of the subcone $C_{\mathbb{G}_f} \equiv \Gamma_5^* \setminus H_f^{(+)} $ are precisely $\mathfrak{g}^- \cup \mathfrak{g}^0$ but this is not the case in general.  There are additional extremal rays for this subcone
which we denote by $\widetilde{\mathfrak{g}}^0$.  Note that $\widetilde{\mathfrak{g}}^0 \subset \Sigma_f$.  The subcone $C_{\mathbb{G}_f}\subset H_f^{(-)}\cup\Sigma_f$ is generated by $\mathbb{G}_f=\{\mathfrak{g}^0,\widetilde{\mathfrak{g}}.^0,\mathfrak{g}^-\}$


By applying the procedure described above, we can obtain the full space of maximin rays $C_{\mathbb{G}_f}^{\text{Mm}}\subset C_{\mathbb{G}_f}$. Repeating the same procedure for all facets of $\Omega_5^*$ we are guaranteed to cover the entirety of its complement in $\Gamma_5^*$. Furthermore, given the large number of symmetries of the cones, we can largely reduce the problem by considering only facets up to permutations.

We have now reduced the problem to the search of all DMP rays inside $29$ different subcones $C_{\mathbb{G}_f}$, one for each of the facets reported in the appendix \ref{sec:appendix}. Due to the large number of generators and the large number of new rays generated by the cutting procedure, for most facets the search is still impractical.  Note however that for a few of the facets the problem can be solved easily and no DMP ray was found outside of $\Omega_5^*$.  
On the other hand, we can take another approach: we only ask whether or not any new maximin provable inequality exists, rather than trying to find the inequalities explicitly. In this case there is an efficient algorithm that we now explain.
The conclusion of implementing this algorithm is that there are no maximin provable inequalities corresponding to rays in $\Gamma_5^*$ outside of $\Omega_5^*$.  For this reason, solving this simpler problem is sufficient.

We now describe this algorithm.  Start by choosing a facet $f$ of $\Omega_5^*$.  From this facet
we can determine the corresponding collection $\mathfrak{g}^-$ of extremal rays of $\Gamma_5^*$ that lie in $H_f^{(-)}$.  Let us denote these generators as
\begin{equation*}
\mathfrak{g}^- = \{ g_1^-, g_1^-, \ldots, g_n^- \}
\end{equation*}
and first consider the  generator $g_1^-$. We want to check efficiently whether or not
\begin{equation}
\label{eq:gen_removal}
C_{\mathbb{G}_f\backslash\{g_1^-\}}^{\mathcal{I}^-}=C_{\mathbb{G}_f}^{\mathcal{I}^- } \qquad \forall\;\mathcal{I}^- \in P.
\end{equation}
If we wanted to explicitly find the cone $C_{\mathbb{G}}^{\mathcal{I}^-}$ using the procedure described above, we would find the extremal rays of the cone defined by the inequalities in \eqref{eq:DMP_condition} in addition to the
conditions $\alpha_i\geq 0$. Instead of solving this problem we can check that the solution is left unchanged under the removal of the generator $g_1^-$, i.e. by setting $\alpha_1^-=0$, where $\alpha_1^-$ is the component of $\alpha$ corresponding to the generator
$g_1^-$. This is equivalent to adding the new constraint $\alpha_1^- \leq 0$ to the system of inequalities. Thus, \eqref{eq:gen_removal} is equivalent to checking that the extra inequality  $\alpha_1^- \leq 0$ is redundant; this redundancy checking can be done efficiently using standard linear programming techniques. 

Suppose that we find that \eqref{eq:gen_removal} is indeed satisfied.  We can then repeat the procedure and ask if

\begin{equation}
\label{eq:gen_removal2}
C_{\mathbb{G}_f\backslash\{g_1^- , g_2^-\}}^{\mathcal{I}^-}=C_{\mathbb{G}_f\backslash\{g_1^- \}}^{\mathcal{I}^-} \qquad \forall\;\mathcal{I}^- \in P.
\end{equation}
This can be done exactly as above except that the space of $\alpha$ parameters is taken to have one less dimension because we put $\alpha_1^- = 0$ in all inequalities. Assuming that
\eqref{eq:gen_removal2} is found to be true, we can continue to iterate this procedure until a DMP inequality is proven to exist outside of  $\Omega_5^*$ or until we conclude that
\begin{equation}
\label{eq:allgenremove}
C_{\mathbb{G}_f\backslash\mathfrak{g}^-}^{\mathcal{I}^-}=C_{\mathbb{G}_f}^{\mathcal{I}^- } \qquad \forall\;\mathcal{I}^-.
\end{equation}
If \eqref{eq:allgenremove} is found to be true, we would conclude that there is no DMP ray in $H_f^{(-)}$. On the other hand, it is still possible that DMP rays exist on $\Sigma_f$, and in fact this must be the case since $\Sigma_f$ also contains the facet $f$ of $\Omega_5^*$, which is generated by DMP rays. However, notice that if, after repeating the same search for all possible choices of $f$, we find that there is no DMP ray in any $H_f^{(-)}$, we can conclude that there is no DMP ray in the entire complement of $\Omega_5^*$ in $\Gamma_5^*$. 

After completing the entire search numerically, we confirm that this is indeed the case. Therefore we conclude that starting from the inequalities of \citep{Bao:2015bfa} (which define $\Gamma_5$), the technique of maximin cannot be used to prove any new useful inequality for dynamical situations.

\section{Discussion}
\label{sec:discussion}

The most important open problem discussed above is to determine whether or not the dynamical holographic entropy cone for $N$ regions
lies within the quantum cone for $N$ regions.   Because so little is known about the quantum cone, we have chosen to trust the
conjecture of \citep{Cadney:2011aa} that all balanced classical inequalities are valid quantum inequalities.  

The immediately startling feature of this assumption is the power of MMI.  The dynamical entropy cone (and its holographic screen generalization \cite{Sanches:2016sxy})
is bounded by the MMI cone, and the MMI cone already satisfies every known balanced classical inequality with the only exception of the family given in  \eqref{eq:zhang}.
For $N=5$ this family furnishes the inequality \eqref{eq:purified_zhang} which cuts the $N=5$ MMI cone.    A proof of \eqref{eq:purified_zhang} for the 
dynamical cone,  while less ambitious than demonstrating that every static inequality holds dynamically, would constitute powerful evidence for the conjecture that
the dynamical entropy cone implies all balanced classical inequalities which, in turn, suggests (by  \citep{Cadney:2011aa}) that the dynamical cone lies in the quantum cone.
We have explained above how this statement motivates the extension of covariant holographic entanglement entropy to general spacetimes.

Unfortunately, our major technical result in section \ref{sec:implications} is that the maximin technique cannot currently prove any inequality that cuts the MMI cone for $N=5$.
This result directly relied on what is presently known about the static holographic entropy cone for five regions.  We expect additional inequalities, beyond those given in appendix
\ref{sec:appendix} to be valid for the static cone.   Upon their discovery our procedure can be performed again, potentially revealing the existence of a nontrivial maximin provable inequality.
We expect that this outcome is unlikely and that new technology is necessary to prove \eqref{eq:purified_zhang} dynamically.

We cannot rule out the possibility that the MMI cone and the dynamical holographic entropy cone coincide.  If this is the case, then there are two possibilities.  If the conjecture of \citep{Cadney:2011aa}
is correct, then we would conclude that certain geometries are not realizable in AdS/CFT, a possibility discussed in  \cite{Marolf:2017shp}.  This would mean that certain cases of the screen entanglement
proposal violate quantum mechanics, suggesting that the construction is unimportant in quantum gravity.  The other possibility is that  the conjecture of \citep{Cadney:2011aa} is false, in which case
we have little guidance with regard to the quantum cone.
Given that our interpretations strongly rely on the assumption that balanced classical inequalities are quantum inequalities,
 it is critical to explore the validity of this conjecture in future work.

%

%

\section{Appendix}
\label{sec:appendix}
 
For the convenience of the reader we report here the new $N=5$ inequalities proven in \cite{Bao:2015bfa} for the static holographic entropy cone.
\begin{widetext}
\begin{center}
\begin{align}
& S(ABC)+S(BCD)+S(CDE)+S(DEA)+S(EAB)\geq\nonumber\\
& S(AB)+S(BC)+S(CD)+S(DE)+S(EA)+S(ABCDE)\label{eq:new_inequality1}\\[8pt]
& 2S(ABC)+S(ABD)+S(ABE)+S(ACD)+S(ADE)+S(BCE)+S(BDE)\geq\nonumber\\
& S(AB)+S(ABCD)+S(ABCE)+S(ABDE)+S(AC)+S(AD)+S(BC)+S(BE)+S(DE)\label{eq:new_inequality2}\\[8pt]
& S(ABE)+S(ABC)+S(ABD)+S(ACD)+S(ACE)+S(ADE)+S(BCE)+S(BDE)+S(CDE)\geq\nonumber\\
& S(AB)+S(ABCE)+S(ABDE)+S(AC)+S(ACDE)+S(AD)+S(BCD)+S(BE)+S(CE)+S(DE)\label{eq:new_inequality3}\\[8pt]
& S(ABC)+S(ABD)+S(ABE)+S(ACD)+S(ACE)+S(BC)+S(DE)\geq\nonumber\\
& S(AB)+S(ABCD)+S(ABCE)+S(AC)+S(ADE)+S(B)+S(C)+S(D)+S(E)\label{eq:new_inequality4}\\[8pt]
& 3S(ABC)+3S(ABD)+3S(ACE)+S(ABE)+S(ACD)+S(ADE)+S(BCD)+S(BCE)+S(BDE)\nonumber\\
& +S(CDE)\geq 2S(AB)+2S(ABCD)+2S(ABCE)+2S(AC)+2S(BD)+2CE)\nonumber\\
& +S(ABDE)+S(ACDE)+S(AD)+S(AE)+S(BC)+S(DE)\label{eq:new_inequality5}
\end{align}
\end{center}
\end{widetext}
Additional (but physically equivalent) inequalities can be found using purifications in the same way as one can obtain weak monotonicity from strong subadditivity.

We also report all of the extremal rays of the $N=5$ MMI cone up to symmetries. We present these rays using the following ordering of the 31 components:
\begin{equation}
\begin{split}
&(A, B, C, D, E, AB, AC, \ldots,  AE, BC , \ldots, DE,\\
  & ABC, \ldots , CDE, ABCD, \ldots, BCDE, ABCDE)
\end{split}
\end{equation}
Here, $A$ means $S(A)$, and so on. 

The first three rays below are inherited from the $N=2,3$, and $4$ cones respectively.  All of the other rays are new. The fifth and sixth rays below are the only ones that violate some permutation of \eqref{eq:purified_zhang}.  

\begin{widetext}
\begin{align}
(1,1,0,0,0,0,1,1,1,1,1,1,0,0,0,0,0,0,1,1,1,1,1,1,0,0,0,0,1,1,0)\nonumber\\
(0,0,1,1,1,0,1,1,1,1,1,1,2,2,2,1,1,1,2,2,2,2,2,2,1,2,2,2,1,1,1)\nonumber\\
(0,1,1,1,1,1,1,1,1,2,2,2,2,2,2,2,2,2,2,2,2,3,3,3,3,3,3,3,3,2,2)\nonumber\\
(1,1,1,1,1,2,2,2,2,2,2,2,2,2,2,2,2,2,2,2,2,2,2,2,2,2,2,2,2,2,1)\nonumber\displaybreak[0]\\
(1,1,1,1,1,2,2,2,2,2,2,2,2,2,2,2,2,2,2,2,2,2,2,2,3,2,2,2,2,2,1)\nonumber\\
(1,1,1,1,1,2,2,2,2,2,2,2,2,2,2,2,2,2,2,2,2,2,2,3,3,2,2,2,2,2,1)\nonumber\\
(1,1,1,1,1,2,2,2,2,2,2,2,2,2,2,2,2,2,2,2,2,2,3,3,3,2,2,2,2,2,1)\nonumber\\
(1,1,1,1,1,2,2,2,2,2,2,2,2,2,2,2,2,2,2,2,3,2,3,2,3,2,2,2,2,2,1)\nonumber\\
(1,1,1,1,1,2,2,2,2,2,2,2,2,2,2,2,2,2,2,2,3,2,3,3,3,2,2,2,2,2,1)\nonumber\\
(1,1,1,1,1,2,2,2,2,2,2,2,2,2,2,2,2,2,2,2,3,3,3,3,3,2,2,2,2,2,1)\nonumber\\
(1,1,1,1,1,2,2,2,2,2,2,2,2,2,2,2,2,2,2,3,3,3,2,3,3,2,2,2,2,2,1)\nonumber\\
(1,1,1,1,1,2,2,2,2,2,2,2,2,2,2,2,2,2,2,3,3,3,3,3,3,2,2,2,2,2,1)\nonumber\\
(1,1,1,1,1,2,2,2,2,2,2,2,2,2,2,2,2,3,3,2,2,3,3,3,3,2,2,2,2,2,1)\nonumber\\
(1,1,1,1,1,2,2,2,2,2,2,2,2,2,2,2,2,2,3,3,3,3,3,3,3,2,2,2,2,2,1)\nonumber\\
(1,1,1,1,1,2,2,2,2,2,2,2,2,2,2,2,2,3,3,2,3,3,3,3,3,2,2,2,2,2,1)\nonumber\\
(1,1,1,1,1,2,2,2,2,2,2,2,2,2,2,1,3,3,3,3,3,3,3,3,3,2,2,2,2,2,1)\nonumber\\
(1,1,1,1,1,2,2,2,2,2,2,2,2,2,2,3,3,3,3,3,3,3,3,3,3,2,2,2,2,2,1)\nonumber\\
(1,1,1,1,2,2,2,2,3,2,2,3,2,3,3,3,3,3,3,3,3,3,3,3,4,4,3,3,3,3,2)\nonumber\\
(1,1,1,1,1,2,2,2,2,2,2,2,2,2,2,3,3,3,3,3,3,3,3,3,3,4,4,4,4,4,3)\nonumber\\
(1,1,1,1,2,2,2,2,3,2,2,3,2,3,3,3,3,2,3,4,4,3,4,4,4,4,3,3,3,3,2)\nonumber\\
(1,1,1,1,2,2,2,2,3,2,2,3,2,3,3,3,3,4,3,4,4,3,4,4,4,4,3,3,3,3,2)\nonumber\\
(1,1,2,2,2,2,3,3,3,3,3,3,4,4,4,4,4,4,3,3,5,5,5,5,4,4,4,4,3,3,2)\nonumber\\
(1,1,1,2,2,2,2,3,3,2,3,3,3,3,4,3,4,4,4,4,5,4,4,5,5,5,5,4,4,4,3)\nonumber\\
(2,2,2,2,3,4,4,4,5,4,4,5,4,5,5,4,6,5,6,5,5,6,5,5,5,6,5,5,5,5,3)\nonumber\\
(2,2,2,2,3,4,4,4,5,4,4,5,4,5,5,4,6,5,6,5,5,6,5,5,7,6,5,5,5,5,3)\nonumber\\
(2,2,2,2,3,4,4,4,5,4,4,5,4,5,5,4,6,5,6,5,5,6,7,5,7,6,5,5,5,5,3)\nonumber\\
(2,2,2,2,3,4,4,4,5,4,4,5,4,5,5,4,6,5,6,5,7,6,7,7,7,6,5,5,5,5,3)\nonumber\\
(2,2,2,2,3,4,4,4,5,4,4,5,4,5,5,4,6,5,6,7,5,6,7,7,7,6,5,5,5,5,3)\nonumber\\
(2,2,2,3,3,4,4,5,5,4,5,5,5,5,6,6,5,7,7,5,6,7,7,6,6,7,7,6,6,6,4)
\label{eq:MMI_rays}
\end{align}
%
\end{widetext}

\begin{acknowledgments}

It is a pleasure to thank  N.~Bao, V.~Hubeny, M.~Rangamani, F.~Sanches, M. Walter and A. Winter for discussions and correspondence. M.R. also would like to thank University College London and the organizers of the workshop ``Gravity - New perspectives from strings and higher dimensions'' at the Centro de Ciencias de Benasque Pedro Pascual, for hospitality during the course of this project.

M.R. was supported by the Simons Foundation via the ``It from Qubit'' collaboration and by funds from the University of California. S.J.W. is supported by the Department of Energy,Office of Science, Office of High Energy Physics, under contract No. DE-SC0011702, and by Foundational Questions Institute grant FQXi-RFP-1507.

\end{acknowledgments}


\bibliographystyle{utcaps}
\bibliography{references}

\end{document}